\documentclass[sigconf]{acmart}

\usepackage{color}
\usepackage{xspace}

\usepackage{adjustbox}
\usepackage{acronym}
\usepackage{multicol}
\usepackage{multirow}
\usepackage{graphicx}
\usepackage{enumitem}

\let\savedegree\bigtimes
\let\bigtimes\relax

\usepackage{mathabx}
\let\bigtimes\savedegree

\acrodef{RS}{Recommender System}
\acrodef{POI}{Point-of-Interest}
\acrodef{CF}{Collaborative Filtering}
\acrodef{GUI}{Graphical User Interface}

\newcommand{\eg}{\textit{e.g., }}
\newcommand{\ie}{\textit{i.e., }}
\newcommand{\etc}{\textit{etc. }}
\newcommand{\yelp}{{Yelp}\xspace}

\newcommand{\gow}{{Gowalla}\xspace}

\newcommand{\pmf}{\texttt{MF}\xspace}
\newcommand{\usg}{\texttt{USG}\xspace}
\newcommand{\frsqr}{{Foursquare}\xspace}
\newcommand{\lore}{\texttt{LORE}\xspace}
\newcommand{\goso}{\texttt{GeoSoCa}\xspace}
\newcommand{\topp}{\texttt{MostPop}\xspace}

\newcommand{\capri}{{\texttt{CAPRI}\xspace}\xspace}

\AtBeginDocument{%
  \providecommand\BibTeX{{%
    \normalfont B\kern-0.5em{\scshape i\kern-0.25em b}\kern-0.8em\TeX}}}

\setcopyright{acmcopyright}
\copyrightyear{2023}
\acmYear{2023}
\acmDOI{XXXXXXX.XXXXXXX}

\acmConference[CIKM '23]{The 32nd ACM International Conference on Information and Knowledge Management}{October 21--25, 2023}{Birmingham, UK}
\acmBooktitle{CIKM '23: The 32nd ACM International Conference on Information and Knowledge Management, October 21--25, 2023, Birmingham, UK} 
\acmPrice{15.00}
\acmISBN{978-1-4503-XXXX-X/18/06}

\begin{document}

\title[CAPRI: Context-Aware Interpretable Point-of-Interest Recommendation Framework]
{CAPRI: Context-Aware Interpretable Point-of-Interest Recommendation Framework}

\author{Ali Tourani}
\email{ali.tourani@uni.lu}
\affiliation{%
  \institution{University of Luxembourg}
  \city{Luxembourg}
  \country{Luxembourg}
}

\author{Hossein A.~Rahmani}
\email{hossein.rahmani.22@ucl.ac.uk}
\affiliation{%
  \institution{University College London}
  \city{London}
  \country{United Kingdom}
}


\author{Mohammadmehdi Naghiaei}
\email{naghiaei@usc.edu}
\affiliation{%
  \institution{University of Southern California}
  \city{California}
  \country{USA}
}

\author{Yashar Deldjoo}
\email{yashar.deldjoo@poliba.it}
\affiliation{%
  \institution{Polytechnic University of Bari}
  \city{Bari}
  \country{Italy}
}

\renewcommand{\shortauthors}{A.~Tourani, H.~A.~Rahmani, M.~Naghiaei, Y.~Deldjoo}

\begin{abstract}

\ac{POI} recommendation systems have gained popularity for their unique ability to suggest geographical destinations, with the incorporation of contextual information such as time, location, and user-item interaction. Existing recommendation frameworks lack the contextual fusion required for \ac{POI} systems. This paper presents \capri, a novel \ac{POI} recommendation framework that effectively integrates context-aware models, such as \goso, \lore, and \usg, and introduces a novel strategy for the efficient merging of contextual information. \capri integrates an evaluation module that expands the evaluation scope beyond accuracy to include novelty, personalization, diversity, and fairness. With an aim to establish a new industry standard for reproducible results in the realm of \ac{POI} recommendation systems, we have made \capri openly accessible on GitHub, facilitating easy access and contribution to the continued development and refinement of this innovative framework.\footnote{\url{https://github.com/CapriRecSys/CAPRI}}

\end{abstract}




\maketitle

\section{Introduction}
\label{sec_intro}
For selecting the appropriate vacation destination, restaurant, visiting locations, and so-called \acf{POI}, users must choose from a variety of possibilities.
\ac{POI} recommender systems can be useful tools for overcoming the inevitable information overload in many use cases.
For instance, recommending hotels and other travel-related destinations remains a challenging task since trip planning entails looking for a set or list of interconnected factors (\eg means of transportation, housing, and attractions) and where contextual factors may have a significant impact (\eg time, location, and social environment).

Recent years have seen the development of numerous frameworks, libraries, and tools for \ac{RS} that make it easy for researchers to mimic the recommendation process and its influence on user preference.
Utilizing frameworks for recommendation leads to the standardization of algorithm implementations and facilitates the reproducibility of experiments.
Despite the advances, reproducibility remains a challenge in \ac{RS} research, particularly in the areas that are not well-established, such as fairness-aware and domain-specific recommendations \cite{sonboli2020fairness}.
Even minor differences in parameters and experimental settings can yield inconsistent results, making it difficult to provide definitive answers about the relative properties of different algorithms.
Hence, reproducible evaluation and fair comparison of methods are demanding factors in \acp{RS}.

Despite the progress made in the field, existing frameworks for the reproducibility of \acp{RS} are typically intended to simulate generic \acl{CF} environments.
For instance, \texttt{Cornac} \cite{salah2020cornac} includes models leveraging auxiliary data such as item descriptive text and image.
\texttt{RecBole} \cite{zhao2021recbole}, an alternative comprehensive framework, introduces general, sequential, and knowledge-based recommendations.
Likewise, \texttt{Elliot} \cite{anelli2021elliot} is another framework that covers a wide range of general-purpose models.

However, \ac{POI} recommendation has particularities that set them apart from recommendations in other domains:

\begin{itemize}
    \item \textbf{The importance of context integration and fusion}:
    The users' check-ins in \ac{POI} recommendations are considerably affected by the contextual information.
    For instance, the geographical property of location affects the user mobility pattern or users' visit is time-depended which indicates the importance of temporal information \cite{rahmani2020joint}.
    Other types of context may include social ties, the category of \acp{POI}, comments on \acp{POI}, \etc
    Previous research works such as \cite{rahmani2022systematic} have shown the way incorporating these \textbf{rich contexts} information have a significant impact on the performance of \ac{POI} recommendation models.
    Therefore, in recent years, there has been a growth in the demand for specialized recommendation algorithms and methodologies that can \textbf{incorporate} and \textbf{fuse contextual} information into the \ac{POI} recommendation process~\cite{werneck2022reproducible};
    
    \item \textbf{High sparsity}:
    The characteristics of the check-in datasets of the \ac{POI} recommendation domain differ significantly from those of the other recommendation domain \cite{bell2007lessons,liu2017exp}.
    Accordingly, the density of \ac{POI} check-ins data is typically approximately 0.1\%, whereas the density of \textit{Netflix} data for movie suggestions is 1.2\%.
    This is because a person can only visit a limited number of locations, whereas a city can contain a vast number of \acp{POI};
    
    \item \textbf{Necessity for multi-dimensional evaluation}:
    Previous papers \citep{liu2017exp,werneck2021systematic,sanchez2021PoInt} in the \ac{POI} field predominantly focus on accuracy-oriented metrics.
    However, there is a remarkable consensus in the \ac{RS} community that there are other important facets to the recommendation process that accuracy metric systems cannot simply capture, such as the novelty, diversity, and catalog coverage of recommenders.
    Therefore, we aim to standardize multi-faceted evaluation on the accuracy, beyond-accuracy, and fairness dimensions \cite{deldjoo2022survey}.
\end{itemize}

\noindent \textbf{Contributions.}
The work at hand addresses the above shortcoming by proposing \textit{\capri}\footnote{\url{https://caprirecsys.github.io/CAPRI/}}, a specialized framework for evaluating and benchmarking state-of-the-art \ac{POI} recommendation models.
Different from existing open-source frameworks, such as \texttt{DaisyRec} \citep{sun2020we}, \texttt{Elliot} \citep{anelli2021elliot}, \texttt{LensKit} \citep{ekstrand2011rethinking}, \texttt{LibRec} \citep{guo2015librec}, \texttt{LibRec-auto} \citep{Sonboli2021}, \texttt{OpenRec} \citep{yang2018openrec}, \texttt{CaseRec} \citep{da2018case}, which mainly aim to reproduce various traditional recommender systems, deep learning-based recommender systems such as \texttt{DeepRec} \citep{zhang2019deeprec}, and multimodal \acp{RS} like \texttt{Cornac} \citep{salah2020cornac}, \capri is intended to provide contextually aware recommendation and evaluation in the \ac{POI} domain.
We have equipped our framework with state-of-the-art models, algorithms, well-known datasets for \ac{POI} recommendations, and multi-dimensional evaluation criteria (accuracy, beyond-accuracy, and user-item fairness).
It also supports the reproducibility of results using various adjustable configurations for comparative experiments.

To the best of our knowledge, there is no publicly accessible framework for the reproducibility of \ac{POI} models in the field of context-aware \ac{POI} recommendation, despite the recent advances in the field.


\section{Related Frameworks}
\label{sec_related}

In recent years, introducing and implementing \Ac{RS} frameworks and libraries gained huge attention.
\citet{Sonboli2021} proposed a recommendation framework titled \texttt{Librec-auto}\xspace for automating various aspects of offline batch \ac{RS} experimentats.
The framework covers a wide range of recommendation and re-ranking algorithms, along with various evaluation and fairness-aware metrics.
Another framework introduced by \citet{Zhao2021} covers a wide range of \ac{RS} applications and contains $73$ models and $28$ datasets.
Their framework, titled \texttt{RecBole}, is implemented in PyTorch and focuses on the performance of the executions, along with covering potential evaluation on the \ac{RS} domain.
\citet{sun2020we} introduced a Python-based toolkit named \texttt{DaisyRec} as a benchmark for rigorous evaluation in recommendation.
Their toolkit is equipped with seven well-tuned state-of-the-art algorithms and six widely-used datasets.
In contrast with other existing open-source libraries, \texttt{DaisyRec} aims to rigorously evaluate the performance of the recommendation.
Similar to \capri, \citet{werneck2022reproducible} introduces an additional framework for the reproducibility of \ac{POI} experiment recommendations.
However, their approach is not exhaustive and is not easily replicable, as it only generates the outcomes of their earlier work \cite{werneck2021systematic}.

The majority of current frameworks are general-purpose and do not prioritize domain-specific recommendation models, such as context awareness.
This characteristic makes it challenging to re-purpose their research skills for domain-specific work.
In contrast to the introduced frameworks, \capri focuses on the \ac{POI} domain and aims to provide researchers with all the necessary resources.

\section{Proposed Framework}
\label{sec_proposed}

\begin{figure*}
    \centering
    \includegraphics[scale=0.3]{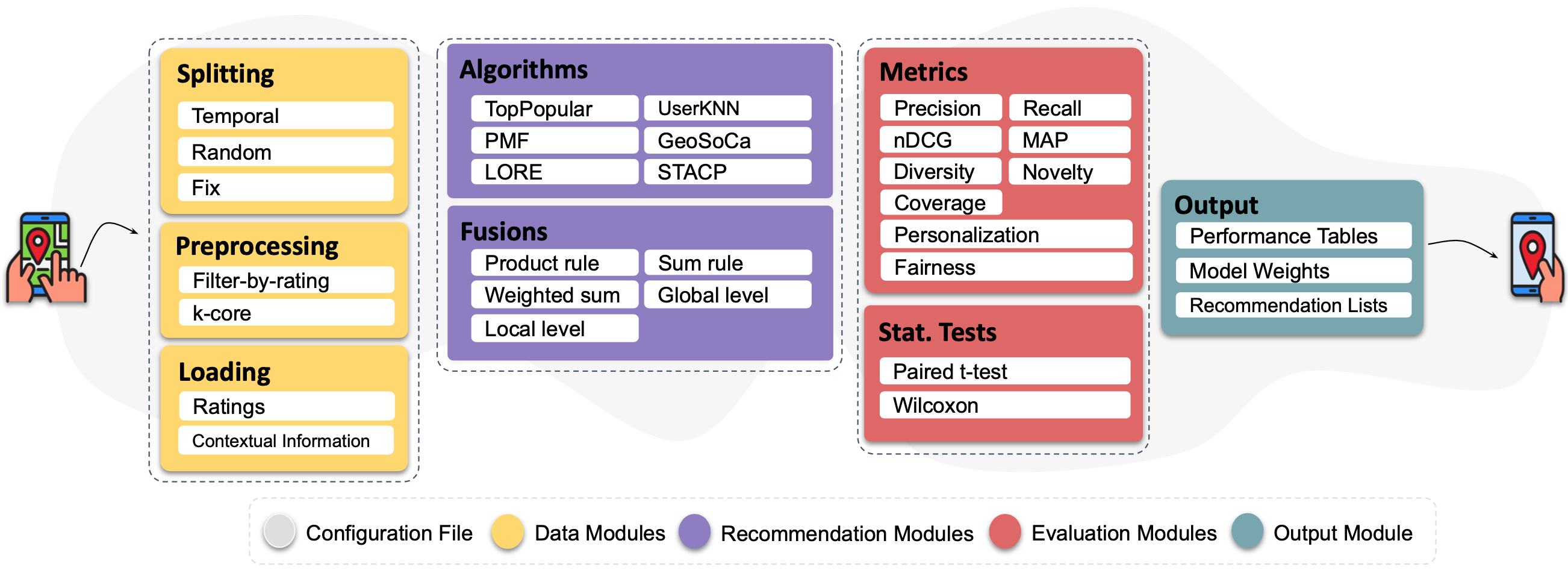}
    \caption{General workflow of the experiments handled by \capri.}
    \label{fig_overview}
\end{figure*}

\capri is an open-source recommendation framework implemented in Python, suitable for practical experimentation and reproducibility study.
The framework is distributed under the \textit{GPL v.3} license and can be downloaded or cloned from GitHub.
In this regard, Figure \ref{fig_overview} illustrates the general workflow of \capri in detail.

\subsection{Files Structure}
\label{sec_file_structure}

In terms of implementation, the files of the framework are organized in several directories to facilitate accessibility and extensibility.
We utilize \textit{PascalCase} and \textit{camelCase} as basic naming structures and merge words into a single string in \capri for folder and file names, respectively.
Detailed descriptions of the directories of the framework containing files are presented below in brief:

\begin{itemize}[leftmargin=*]
    \item \textbf{Data:}
    Contains data-driven files and functions of various types.
    Each dataset includes files with the \texttt{.txt} extension that contain train, test, and tune data.
    Moreover, other files containing the check-ins data and relations among users/items, such as social and geographical data, are stored in folders with the same name as each dataset.
    There are also some data processing functions in the Data directory, including \texttt{readDataSizes.py} to read meta-data of the dataset, \texttt{loadDatasetFiles.py} to load selected dataset items, and \texttt{calculateActiveUsers.py} to calculate Active/Inactive users of a selected dataset for fairness-aware analysis.
    Current datasets of \capri will be discussed in Section \ref{sec_datasets}.
    
    \item \textbf{Models:}
    Contains the models used in the framework and several common functions in the \texttt{utils.py} file to avoid code duplication and increase the re-usability of model files.
    For each model, there is a folder with the same name, a \texttt{main.py} to control the overall processing of functions, and varying processing functions to process a selected dataset according to the selected model.
    The accessible models in \capri will be discussed in more depth in Section \ref{sec_models}.
    
    \item \textbf{Evaluations:}
    Contains all evaluation metrics available for analyzing the performance of models on datasets, the evaluator function \texttt{evaluator.py} that leverages the metrics, and a unit test file \texttt{test.py} for evaluating the performance of each measure with different input types.
    The evaluation metrics supplied in \capri, as well as the evaluation process, will be discussed in detail in Sections~\ref{sec_evals} and ~\ref{sec_eval_process}, respectively.
    
    \item \textbf{Outputs:}
    \capri stores the final findings, including ranked lists and evaluation outputs, for reproducibility purposes.
    The file naming structure prohibits a previously performed analysis from being reprocessed.
    It should be noted that due to the size of the ranked list files, we do not save them on GitHub.
\end{itemize}

\noindent Regarding the fact that \capri is open source and accepts contributions, it is straightforward to add new models, datasets, and metrics.
We welcome academics and developers who aim to contribute to the framework's enhancement.
Consequently, documentation on how to contribute to the project is available at the \textit{readthedocs} page of the framework\footnote{\url{https://capri.readthedocs.io/en/latest/}}.




\subsection{Datasets}
\label{sec_datasets}

In the current version of the framework, we have provided modified versions of three popular check-ins datasets: \gow\footnote{\url{http://www.gowalla.com/}}, \yelp\footnote{\url{https://www.yelp.com/dataset}} \cite{liu2017exp}, and \frsqr\footnote{\url{https://sites.google.com/site/yangdingqi/home/foursquare-dataset}}.
The characteristics of the mentioned datasets are presented in Table~\ref{tbl_datasets}.

\begin{table}
  \caption{Characteristics of the datasets available in \capri.}
  \centering
  \label{tbl_datasets}
  \begin{adjustbox}{max width=\columnwidth}
  \begin{tabular}{lccccc}
    \toprule
    \small
    \textbf{Dataset} & \#users & \#POIs & \#check-ins & \#social & \#category \\
    \midrule
    \multirow{1}{*}{\textbf{\yelp}} & 7,135 & 15,575 & 299,327 & 46,778 & 582 \\
    \midrule
    \multirow{1}{*}{\textbf{\gow}} & 5,628 & 30,943 & 618,621 & 46,001 & $-$ \\
    \midrule
    \multirow{1}{*}{\textbf{\frsqr}} & 7,642 & 28,483 & 512,532 & $-$ & $-$ \\
    \midrule
    \bottomrule 
  \end{tabular}
  \end{adjustbox}
\end{table}

\subsection{Models}
\label{sec_models}

\capri covers the recent implementations of various models, which can be applied to the introduced datasets for evaluation and reproducibility goals.
The models implemented in this framework are listed below:

\begin{itemize}[leftmargin=*]
    \item \textbf{\goso:}
    As introduced in \citet{Zhang2015}, this model covers geographical, social, and categorical correlations among users and \acp{POI}.
    These contexts are learned using users' historical check-in data to produce relevance scores for unseen locations.
    \item \textbf{\lore:}
    Another model utilized in \capri is \lore \citet{Zhang2014LORE}, a popular and robust model for location recommendation focused on the impacts of geographical and social influence on users' check-in behaviors.
    \item \textbf{\usg:}
    As introduced in \citet{Ye2011}, \usg takes geographical influence, social network, and user interest into account for \ac{POI} recommendation.
\end{itemize}

The current \capri version covers standard competitive contextual models for the \ac{POI} domain, with users having the flexibility to modify contexts per their requirements. Our future plans include the incorporation of deep learning, graph-based, sequential, and sessions-based models as proposed in works like \cite{adamczak2020session, quadrana2017personalizing}. These models can integrate various contextual components like geographical, temporal, social, and categorical relevance scores using fusion rules such as \textit{product} or \textit{sum}~\cite{liu2017exp,rahmani2022systematic}, forming a unified preference score \cite{rahmani2019lglmf,rahmani2019category,rahmani2022exploring}. Contextual information, denoted by $c_i$, can be infused using a polynomial regression model

\begin{small}
\begin{equation}
rec_{u,p} = \mathbf{\Lambda} \cdot \mathbf{C} + \mathbf{\Lambda}_{\text{pair}} \cdot \mathbf{C}_{\text{pair}} + \lambda_{123}c_1c_2c_3
\label{eq:ployregcontext}
\end{equation}
\end{small}
where:
\begin{itemize}
\item $\mathbf{\Lambda} = [\lambda_1, \lambda_2, \lambda_3]$
\item $\mathbf{C} = [c_1, c_2, c_3]$
\item $\mathbf{\Lambda}_{\text{pair}} = [\lambda_{12}, \lambda_{13}, \lambda_{23}]$
\item $\mathbf{C}_{\text{pair}} = [c_1c_2, c_1c_3, c_2c_3]$
\end{itemize}


\noindent in which $\lambda_j$ indicates the importance weight for the context $c_i$ learned by the model. Note that the product rule ($\bigodot$) would have $\lambda_j =0$ for all $j$ and $\lambda_{123}=1$. In the case of the sum ($\bigoplus$), $\lambda_1$, $\lambda_2$, and $\lambda_3$ are $1$ and the rest are $0$, while in the weighted sum ($\bigboxplus$), optimal values are assigned to them to maximize performance criteria.

\begin{table*}[!htbp]
    \caption{Accuracy and beyond-accuracy performance of models and baselines evaluated on top-20 recommendation lists on the \yelp dataset. In context, g, t, s, and c show the geographical, temporal, social, and categorical contexts, respectively. (\textbf{Bold} the best metric values).}
    \label{tbl_result}
    \begin{adjustbox}{max width=\textwidth}
        \begin{tabular}{llllllllllccccc}
            \toprule
            \multirow{2}{*}{\textbf{Model}} & \multicolumn{4}{c}{\textbf{Context}} & \multirow{2}{*}{\textbf{Fusion}} & \multicolumn{3}{c}{\textbf{Accuracy}} && \multicolumn{3}{c}{\textbf{Beyond-accuracy}} && \textbf{Fairness} \\
            \cmidrule{2-5}
            \cmidrule{7-9}\cmidrule{11-13}\cmidrule{15-15}
            & g & t & s & c && \multicolumn{1}{c}{$nDCG$} & \multicolumn{1}{c}{$Precision$} & \multicolumn{1}{c}{$Recall$} && $Coverage$ & $Novelty$ & $Diversity$ && $MADr$ \\
            \midrule
            \topp & - & - & - & - & - & 0.0073 & 0.0064 & 0.0185 && 0.12 & 4.9590 & 0.6938 && \textbf{0.0152}\\
            \pmf  & - & - & - & - & - & 0.0078 & 0.0070 & 0.0181 && 0.3 & 4.9736 & 0.6877 && 0.0219 \\
            \midrule
            \multirow{3}{*}{\goso} & \multirow{3}{*}{$\checkmark$} &\multirow{3}{*}{$\times$} & \multirow{3}{*}{$\checkmark$} & \multirow{3}{*}{$\checkmark$} & $\bigodot$ & 0.0169 & 0.0134 & 0.0341 && 49.4 & 7.9083 & 0.6877 && 0.0443 \\
            &&  &  &  & $\bigoplus$ & 0.0160 & 0.0138 & 0.040 && 73.26 & \textbf{8.5047} & 0.7166 && 0.0525 \\
            &&  &  &  & $\bigboxplus$ & 0.0170 & \textbf{0.0145} & \textbf{0.0419} && 69.75 & 8.3311 & 0.6835 && 0.0465 \\
            \midrule
            \multirow{3}{*}{\lore} & \multirow{3}{*}{$\checkmark$} & \multirow{3}{*}{$\checkmark$} & \multirow{3}{*}{$\checkmark$} & \multirow{3}{*}{$\times$} & $\bigodot$ & \textbf{0.0174} & 0.0139 & 0.0335 && 39.14 & 7.7314 & 0.7509 && 0.0530 \\
            &  &  &  &  & $\bigoplus$ & 0.0155 & 0.0134 & 0.0388 && \textbf{74.97} & 8.4723 & \textbf{0.8065} && 0.0404 \\
            &  &  &  &  & $\bigboxplus$ & 0.0169 & 0.0143 & 0.0412 && 74.51 & 8.1677 & 0.7971 && 0.0404 \\
            \midrule
            \bottomrule
        \end{tabular}
    \end{adjustbox}
\end{table*}

\subsection{Evaluation Dimensions}
\label{sec_evals}

\capri is highly compatible with a range of evaluation metrics.
Accordingly, the evaluation metrics available in the framework can be classified into the following categories:

\begin{itemize}[leftmargin=*]
    \item \textbf{Accuracy:}
    for accuracy evaluation, the framework covers \textit{Precision@k}, \textit{Recall@k}, \textit{mAP@k}, and \textit{nDCG@k} metrics, in which $k$ represents the number of items filtered for recommendation.
    
    \item \textbf{Beyond-Accuracy:}
    this category contains \textit{List Diversity}, \textit{Novelty}, \textit{Catalog Coverage}, and \textit{Personalization} metrics.
    
    \item \textbf{Fairness:}
    it contains modules for grouping users and items according to a sensitive attribute.
    Thus, it includes \textit{MADr} and \textit{GCE} evaluation among the user/item groups \cite{deldjoo2021flexible,deldjoo2022survey,rahmani2022unfairness}.
\end{itemize}

It is observable that the offered metrics are tailored to meet the recommendations of \ac{POI} recommendation.
All the evaluation metrics can be accessed using the \texttt{Evaluations} directory in the framework.
There is also a \texttt{test.py} file in the same folder for evaluating the performance of each metric through unit testing.

\subsection{Configuration}
\label{sec_config}

\capri contains a \texttt{config.py} file that provides adjustments and configurations for running various experiments.
Accordingly, the parameters that can be set using the mentioned file are listed below:

\begin{itemize}
    \item {\verb|dataDirectory|}: the path from which the dataset files are read,
    \item {\verb|outputsDir|}: the path to store final recommendation lists generated by the framework,
    \item {\verb|topK|}: \textit{Top-k} items for doing the evaluations (default: 10),
    \item {\verb|limitUsers|}: limiting the number of users in the dataset (default: -1),
    \item {\verb|listLimit|}: limiting the length of the final recommendation lists (default: 10),
    \item {\verb|activeUsersPercentage|}: calculating a list of pre-defined groups of users known as \textit{"active users,"}
    \item {\verb|models|}: available models to be selected by the user,
    \item {\verb|datasets|}: available datasets to be selected by the user,
    \item {\verb|fusions|}: available fusions to be selected by the user,
    \item {\verb|evaluationMetrics|}: available evaluation metrics to be selected by the user.
\end{itemize}

\section{Evaluation Process}
\label{sec_eval_process}
This section describes how the evaluation process takes place in \capri.
All the evaluation-related functions of the framework are collected in \texttt{evaluator.py} file in the \textit{Evaluations} directory.
Accordingly, the model and the dataset selected by the user, as well as the evaluation and model parameters, are passed to the evaluator.
Model parameters contain model-specific final scores, such as geographical, social, and categorical calculations for \goso.
Evaluation parameters are formed as a dictionary that contains evaluation-related feed, such as the list of users, the list of \acs{POI}, and ground truth data.
By iterating over users in the dataset, overall recommendation scores and requested accuracy measures such as \textit{Precision@k} and \textit{Recall@k} are calculated.
The final results will be saved in files for later processes.

\section{Benchmarking}
\label{sec_benchmark}

Table~\ref{tbl_result} shows initial and experimental results using our proposed framework, \capri.
As one can see, using \capri, we are able to incorporate different contextual models of the \ac{POI} recommendation domain as well as different approaches to context fusion.
We can see that the fusion methods have a great impact on the performance of \ac{POI} models.
The \textit{sum} rule could show a much better impact on the beyond-accuracy performances (\ie coverage, novelty, and diversity) compared to the \textit{product} rule.
Additionally, it can be seen that the \textit{sum} rule often outperforms the \textit{product} operation in accuracy for the \goso model.
The \textit{product} operation often outperforms the \textit{sum} operation in accuracy for \lore (with the exception of $Recall$ where the \textit{sum} is better than the \textit{product}).
Finally, the \textit{weighted sum} has a favorable effect on the \goso model, making it the most accurate model and enhancing the \textit{sum} model under \lore. As a work where \capri has been employed for \ac{POI} recommendation, further evaluation of the results for user/item fairness can be found in \cite{rahmani2022rule}.

\section{Discussion and Conclusion}
\label{sec_conclusion}

The open-source framework, \capri, developed for \ac{POI} recommendation systems, distinguishes itself by leveraging contextual information in the suggestion pipeline. It incorporates robust models, datasets, and evaluation metrics to provide proper location suggestions matching the context. Like any software, constant development is necessary for \capri to accommodate user needs. We plan on widening its coverage of datasets and models, incorporating more evaluation metrics, and investigating parallelization and requests queuing to improve performance. Future iterations will support batch request handling with user-defined parameters and offer a GUI for easier configuration. We aim to make it directly installable via Python's \texttt{pip}, and integrate bias mitigation approaches to reduce system biases. The framework is publicly available on GitHub for researchers.

\bibliographystyle{ACM-Reference-Format}
\bibliography{references}


\end{document}